\documentclass[twocolumn]{article}    

\def\be{\begin{equation}}
\def\ee{\end{equation}}
\def\bea{\begin{eqnarray}}
\def\eea{\end{eqnarray}}

\usepackage{graphicx}
\usepackage{dcolumn}
\usepackage{bm}
\usepackage{tikz}
\usepackage{pgfplots}
\usepackage[colorinlistoftodos]{todonotes}
\pgfplotsset{width=10cm,compat=1.9}
\usepackage{physics}
\usepackage{caption}
\usepackage{yfonts}
\usepackage{euscript}
\usepackage{subcaption}
\usepackage{multirow}
\usepackage{caption} 
\usepackage{booktabs}
\usepackage{kantlipsum,cuted}
\usepackage{mathtools}
\usepackage{amsmath}
\usepackage{amssymb}
\usepackage{multirow}
\usepackage{amsthm}
\usepackage{authblk}
\newcommand{\unit}{1\!\!1}

\newcommand{\loopdef}{
\begin{tikzpicture}[scale=0.9,baseline=0.3ex]
\draw (0,0) -- (0.3,0) -- (0.3,0.3) -- (0,0.3) -- (0,0);
\node [circle,fill,inner sep=0.75] at (0,0) {};
\node [circle,fill,inner sep=0.75] at (0.15,0) {};
\node [circle,fill,inner sep=0.75] at (0.3,0) {};
\node [circle,fill,inner sep=0.75] at (0,0.3) {};
\node [circle,fill,inner sep=0.75] at (0.15,0.3) {};
\node [circle,fill,inner sep=0.75] at (0.3,0.3) {};
\node [circle,fill,inner sep=0.75] at (0,0.15) {};
\node [circle,fill,inner sep=0.75] at (0.3,0.15) {};
\end{tikzpicture}}

\newcommand{\vacedge}{
\begin{tikzpicture}[scale=0.9,baseline=0.3ex]
\draw (0,0) -- (0.3,0) -- (0.3,0.3) -- (0,0.3) -- (0,0);
\draw (0,0.15) -- (0.15,0.15) -- (0.3,0.15);
\draw (0.15,0.0) -- (0.15,0.15) -- (0.15,0.3);
\node [circle,fill,inner sep=0.75] at (0,0) {};
\node [circle,fill,inner sep=0.75] at (0.15,0) {};
\node [circle,fill,inner sep=0.75] at (0.3,0) {};
\node [circle,fill,inner sep=0.75] at (0,0.3) {};
\node [circle,fill,inner sep=0.75] at (0.15,0.3) {};
\node [circle,fill,inner sep=0.75] at (0.3,0.3) {};
\node [circle,fill,inner sep=0.75] at (0,0.15) {};
\node [circle,fill,inner sep=0.75] at (0.3,0.15) {};
\node [circle,fill,inner sep=0.75] at (0.15,0.15) {};
\end{tikzpicture}}

\newcommand{\oneedge}{
\begin{tikzpicture}[scale=0.9,baseline=0.3ex]
\draw (0,0) -- (0.3,0) -- (0.3,0.3) -- (0,0.3) -- (0,0);
\draw (0.15,0.15) -- (0.15,0.3);
\node [circle,fill,inner sep=0.75] at (0,0) {};
\node [circle,fill,inner sep=0.75] at (0.15,0) {};
\node [circle,fill,inner sep=0.75] at (0.3,0) {};
\node [circle,fill,inner sep=0.75] at (0,0.3) {};
\node [circle,fill,inner sep=0.75] at (0.15,0.3) {};
\node [circle,fill,inner sep=0.75] at (0.3,0.3) {};
\node [circle,fill,inner sep=0.75] at (0,0.15) {};
\node [circle,fill,inner sep=0.75] at (0.3,0.15) {};
\node [circle,fill,inner sep=0.75] at (0.15,0.15) {};
\end{tikzpicture}}

\newcommand{\twoedge}{
\begin{tikzpicture}[scale=0.9,baseline=0.3ex]
\draw (0,0) -- (0.3,0) -- (0.3,0.3) -- (0,0.3) -- (0,0);
\draw (0.15,0.15) -- (0.15,0.3);
\draw (0.15,0.15) -- (0.3,0.15);
\node [circle,fill,inner sep=0.75] at (0,0) {};
\node [circle,fill,inner sep=0.75] at (0.15,0) {};
\node [circle,fill,inner sep=0.75] at (0.3,0) {};
\node [circle,fill,inner sep=0.75] at (0,0.3) {};
\node [circle,fill,inner sep=0.75] at (0.15,0.3) {};
\node [circle,fill,inner sep=0.75] at (0.3,0.3) {};
\node [circle,fill,inner sep=0.75] at (0,0.15) {};
\node [circle,fill,inner sep=0.75] at (0.3,0.15) {};
\node [circle,fill,inner sep=0.75] at (0.15,0.15) {};
\end{tikzpicture}}

\newcommand{\threeedge}{
\begin{tikzpicture}[scale=0.9,baseline=0.3ex]
\draw (0,0) -- (0.3,0) -- (0.3,0.3) -- (0,0.3) -- (0,0);
\draw (0.15,0.15) -- (0.3,0.15);
\draw (0.15,0.0) -- (0.15,0.15) -- (0.15,0.3);
\node [circle,fill,inner sep=0.75] at (0,0) {};
\node [circle,fill,inner sep=0.75] at (0.15,0) {};
\node [circle,fill,inner sep=0.75] at (0.3,0) {};
\node [circle,fill,inner sep=0.75] at (0,0.3) {};
\node [circle,fill,inner sep=0.75] at (0.15,0.3) {};
\node [circle,fill,inner sep=0.75] at (0.3,0.3) {};
\node [circle,fill,inner sep=0.75] at (0,0.15) {};
\node [circle,fill,inner sep=0.75] at (0.3,0.15) {};
\node [circle,fill,inner sep=0.75] at (0.15,0.15) {};
\end{tikzpicture}}

\title{Quantum Mechanics and the Continuum Limit of an Emergent Geometry}
\begin{document}
\author[1,2]{Philip Tee}
\affil[1]{The Beyond Center for Fundamental Science, Arizona State University, Tempe AZ}
\affil[2]{Department of Informatics, The University of Sussex, Falmer, Brighton. UK}

\maketitle

\begin{abstract}
Recent advances in emergent geometry have identified a new class of models that represent spacetime as the graph obtained as the ground state of interacting Ising spins.
These models have many desirable features, including stable excitations possessing many of the characteristics of a quantum particle.
We analyze the dynamics of such excitations, including a detailed treatment of the edge states not  previously addressed.
Using a minimal prescription for the interaction of defects we numerically investigate approximate bounds to the speed of propagation of such a `particle'.
We discover, using numerical simulations, that there may be a Lieb-Robinson bound to propagation that could point the way to how a causal structure could be accommodated in this class of emergent geometry models.
\end{abstract}

\section{Introduction}
\subsection{Background and motivation}
The search for a consistent theory of Quantum Gravity (QG) has so far not produced a finite and self consistent theory.
Of the many attempts to create a consistent formalism, we can identify two broad classes of approach.
The first class work within a continuous framework, and we include in that semi-classical approaches to QG \cite{birrell1984quantum}, the canonical formalisms of the ADM theory \cite{arnowitt1959quantum}, and of course String Theory \cite{horowitz2005spacetime}.
These approaches do not address the quantization of spacetime itself.
Instead, it is assumed that the physics described by these theories emerge from a mathematical formalism based upon a  preexisting and continuous structure of space and time that exists external to the theory.
In short physics happens {\sl in} space and time.

The second class of theories directly addresses the quantization of spacetime, and does not assume that spacetime is either pre-existent or continuous.
This class of theories includes Quantum Graphity \cite{konopka2006quantum}, Causal Set theory \cite{dowker2006causal}, Dynamical Triangulations \cite{ambjorn2013euclidian}, and Loop Quantum Gravity \cite{rovelli2014covariant,smolin2004invitation}.
Finally and more recently, Wolfram {\sl et al} proposed an entirely abstract method of emerging a discrete spacetime based upon `branchial graphs' \cite{gorard2020some}.
Although entirely different from the models described here it shares the commonality of representing emergent geometry as a graph.
It should also be remarked that there are `matrix model' formulations of type IIB String theories \cite{ishibashi1997large,klinkhamer2020iib} based upon a Lagrangian with an $SU(N)$ gauge symmetry where $N$ is very large.
In these matrix models, spacetime `emerges', with Lorentzian signature metrics, by identification of the eigenvalues of the $N \times N$ gauge field matrices with spacetime points.
As $N \rightarrow \infty$, smooth spacetime is recovered.
In all of these models, physics happens {\sl along with} space and time.

For this second class of models spacetime is effectively discrete at the very short range.
Although there is no hard evidence for this discreteness, it is widely accepted that there exists a length below which measurement makes no sense \cite{hossenfelder2013minimal}.
The model that we focus upon in this work is broadly part of the Combinatorial Quantum Gravity (CQG) program, and starts with the view  that spacetime is fundamentally discrete.
In particular, one such model of an emergent geometry has been the subject of recent study, originating from an original proposal of Trugenberger that describes the emergence of geometry as a phase transition in a `hot soup' of entangled qubits \cite{trugenberger2015quantum}.
This work was  subsequently extended  \cite{trugenberger2016random,trugenberger2017combinatorial,kelly2019self} and used to develop the emerged geometry as a basis for a combinatorial approach to gravity.
The nature of the model is a double Ising interaction defined separately for the vertices and edges of a graph.
It is to be understood that the vertices represent points in spacetime, and the edges the locality relationship between those points.
As the edges have finite length (one implicitly assumes this to be the Planck length $l_p$, but in principle it is any small finite length which captures the discreteness of the model).
The Ising interactions operate in opposition, balancing the alignment of spins at the vertices connected by an edge, against an opposing term that frustrates, by means of an energy penalty, the creation of  edges.
We refer to this type of  model and derivatives as `Ising emergent geometries'.
A less well studied aspect of the model is that it admits stable defects in the geometry.
These defects posses some of the properties of quantized particle that exhibits quantum dynamics in the continuum limit \cite{tee2020dynamics}.

The importance of dynamics is closely linked to the problem of incorporating a causal structure into the emerged geometry, and a well formed explanation of how time emerges to posses distinct features from the other spatial dimensions in the geometry.
Emerging such a `temporal' dimension is a non-trivial consideration.
It has been speculated \cite{trugenberger2016random,tee2020dynamics} that the persistent discrepancy between extrinsic and intrinsic dimension of the ground state graphs (with extrinsic dimension remaining  higher than intrinsic dimension) may indicate the one of the graph's dimensions being spatially inaccessible, and therefore temporal.

The subject of this work is to explore the dynamics of the stable defects of the model, and in particular what they might reveal regarding causal structure.
For the defects to exhibit dynamics a minimal prescription for the interaction of a defect with neighboring vertices is required.
The interaction should be constrained by considerations of locality and consistency, and locality further strengthened by an inverse distance dependence of the interaction.
It is already well known that interacting quantum systems, defined upon a lattice, exhibit approximate causality in the form of a Lieb-Robinson (LR) bound \cite{lieb1972finite}, which limits the speed that particular forms of interaction can propagate in the system.
In particular the interactions need to be long range in nature, and it has been recently shown \cite{tran2020hierarchy} that power law interactions produce LR bounds.
The presence of such a bound amounts to a restricted and approximate form of causality.

We begin in Section \ref{sec:qmd} with a brief overview of the Ising emergent geometry models, including a discussion of the emergence of the stable defects we believe are a candidate for modeling matter and dynamics.

We have previously proposed a minimal form of interaction for stable defects in Ising emergent geometries \cite{tee2020dynamics}, and the argument used to propose the interaction introduced a power law dependence on the strength of it.
In the original treatment the interaction Hamiltonian was a simplified one that only dealt with one half of the Hilbert space used to define the Ising model.
Specifically the defect was treated as simply a vertex phenomenon, and the behavior of the spin states on the edges was not dealt with rigorously.
In Section \ref{sec:dynamics} we extend the treatment to include the edge states and propose a complete form of the interaction Hamiltonian in Eq. \eqref{eqn:dynamic}.
We proceed to show that the correspondence with non-relativistic dynamics in the continuum limit still holds once the edge states are incorporated, and discuss the details of how the interaction Hamiltonian can cause a defects interaction with remote vertices to propagate without recourse to the continuum limit.
This is the first contribution of this work.

We describe in Section \ref{ssec:lieb} how this form of interaction could admit a LR bound.
At this point it is not possible to analytically compute the magnitude of the bound, but we can explore numerically the behavior of the interaction on a single defect in an emerged ground state.
We describe these results in Section \ref{sec:simulations}.
To undertake these simulations we compute the time evolution of the state of the vertex spin states for a graph containing a defect.
This evolution is subject to the interaction Hamiltonian we describe in Section \ref{sec:dynamics}, and we calculate the evolution of the defect subject to this interaction in the discrete model.
We discover that the simulation does indicate an approximate bound on the speed of propagation, and this is the second contribution of this work.
This is an intriguing result as the incorporation of causality in the Ising emergent geometry models is not immediately evident, and is an important open problem.

We conclude with a brief summary of the findings in Section \ref{sec:conclusion}.

\section{Ising Emergent Geometries}
\label{sec:qmd}
\subsection{Ground state model and definitions}
The Ising models of emergent spacetime, originally proposed by Trugenberger \cite{trugenberger2015quantum} and extended by him and others \cite{trugenberger2016random,trugenberger2017combinatorial,tee2020dynamics}, are formulated as an interaction model of spin `qubits' located on the vertices of a graph.
Edges are established by interactions  between spins that align due to their ferromagnetic interaction.
To prevent a condensate creating a perfect graph, operating in opposition to edge formation is a link frustration term, expressed as an anti-ferromagnetic interaction between spin states defined on the edges.

The detail of the model involves defining the Hilbert spaces for the edges and the $N$ vertices of a simple, undirected graph $G(V,E)$ where $V$ is the set of vertices and $E \subset \{V \times V \}$ the edges connecting any two distinct vertices.
At each vertex $v_i $ we define the following Hilbert space $\mathcal{H}_{i} = \mbox{span} \{ \ket{i,0}, \ket{i,1} \}$, and define a spin operator obeying $\hat{s}_{i} \ket{i,s}=s_i\ket{i,s}$ on it.
On this space we impose fermionic anti-commutator algebra,
\begin{align}
	\{ \hat{s}^{+}_i, \hat{s}^{+}_j \},  \{ \hat{s}^{-}_i, \hat{s}^{-}_j \} = 0, \{ \hat{s}^{-}_i, \hat{s}^{+}_j \}&= \delta_{ij} \mbox{,} \\
	\hat{s}^{+}_i \ket{ i,1} = 0, \hat{s}^{+}_i \ket{ i,0} &= \ket{i,1} \mbox{,} \\
	\hat{s}^{-}_i \ket{ i,1} = \ket{i,0}, \hat{s}^{-}_i \ket{ i,0} &= 0 \mbox{.}
\end{align}

At the edges we have $\frac{1}{2}N(N-1)$ Hilbert spaces $\mathcal{H}_{ij} = \mbox{span} \{ \ket{i,j,0}, \ket{i,j,1} \}$.
We interpret the state $\ket{i,j,1}$ as the presence of an edge, and $\ket{i,j,0}$ its absence.
To distinguish the edges from the vertices, we propose edge creation $\hat{a}_{ij}^{\dagger}$  and edge annihilation $\hat{a}_{ij}$ operators.
These operators have the following fermionic algebra ,

\begin{align}
	\{ \hat{a}^{\dagger}_{ij}, \hat{a}^{\dagger}_{kl} \},  \{ \hat{a}_{ij}, \hat{a}_{kl} \} = 0, \{ \hat{a}_{ij}, \hat{a}^{\dagger}_{kl} \}&= \delta_{ik} \delta_{jl} \mbox{,}\\
	\hat{a}^{\dagger}_{ij} \ket{ i, j, 1} = 0, \hat{a}^{\dagger}_{ij} \ket{ i, j, 0} &= \ket{i, j, 1} \mbox{,}\\
	\hat{a}_{ij} \ket{ i, j, 1} = \ket{i,j, 0}, \hat{a}_{ij} \ket{ i, j, 0} &= 0 \mbox{.}
\end{align}

A graph can be completely defined by its adjacency matrix $A_{ij}$, in which a non zero value of $A_{ij}$ indicates the presence of an edge between $v_i$ and $v_j$, with zero elsewhere and on the diagonal (as the graph is simple and has no self-loops).
It will be useful later to note that this matrix can be represented using these edge annihilation/creation operators as $A_{ij} = \hat{a}^{\dagger}_{ij} \hat{a}_{ij}$.
To complete the model we propose a Hamiltonian, with the ground state graph representing the interaction and matter free vacuum of the model.
The original model,termed a Dynamical Graph Model (DGM), used the following elegant and simple Hamiltonian with a dimensionless coupling constant $g$ \cite{trugenberger2015quantum}, 

\begin{equation}\label{eqn:trugenberger}
	H_{DGM}=\frac{g^2}{2} \Bigg ( \sum\limits_{i \neq j}^N \sum\limits_{ k \neq i,j }^N A_{ik}A_{kj}  \Bigg ) - \frac{g}{2}  \sum\limits_{i,j} s_i A_{ij}  s_j \mbox{.}
\end{equation}

This model however possessed an excess of clustering, and a physical vacuum needs high locality which necessitates the absence of clustering.
To cure this, we can exploit the property of the adjacency matrix that the $n^{th}$ power of it counts the number of $n$ length paths between vertices in the graph.
As such adding in a term in $A^3_{ij}$ can be used to suppress triangles, and therefore clustering \cite{trugenberger2016random,tee2020dynamics}.
This model is referred to as Quantum Mesh Dynamics (QMD), as it formed the basis of a dynamical theory of matter in the emerged ground state.
The Hamiltonian proposed in this model is,

\begin{equation}\label{eqn:qmd_hamiltonian}
	H_{QMD}=\frac{g^2}{2} \Bigg ( \Tr A^3 +  \sum\limits_{i \neq j}^N \sum\limits_{ k \neq i,j }^N A_{ik}A_{kj}  \Bigg ) - \frac{g}{2}  \sum\limits_{i,j} s_i A_{ij}  s_j \mbox{.}
\end{equation}

Solving numerically for the ground states of this Hamiltonian at different values of $g$, it was found that they possess certain attractive features:

\begin{description}
    \item [Regular, Euclidean flat ground state] The ground state corresponds to a regular graph where nearly all of the nodes posses the same average degree (number of incident edges) $k$. This configuration is referred to as a `large world' in network science, which means that the graph has a high degree of locality with few `short-cuts' between distant nodes.
    \item [Low dimensionality] There are several ways to quantify the dimension of a graph, both intrinsic dimension (as would be measured by an observer confined to the graph), and extrinsic dimension measured from the point of view of a higher dimensional space in which the graph is embedded \cite{ambjorn1997quantum}. These two measures agree for the ground state until around $d=4$, at which point the extrinsic dimension does not reduce any further. This points to a `preferred' low dimension for the graph.
    \item [Entropy Area Law] It is possible to demonstrate that the ground state graph possesses a measure of informational entropy which is related to the size of the boundary of the graph or any defects in it, rather than its bulk.
\end{description}

An important unanswered question with this and other similar models is the role of time and causality.
As it stands the model does not have a dimension that is intrinsically temporal.
For the purposes of our simulations later, we take the view that time is a label on the states of the model, which can be continuous or discrete.
We refer to this time as `clock time', that is the external time governing dynamical evolution.
It is possible that this definition could be abandoned if more was understood regarding the nature of time in the Ising models, but for the purposes of our investigation we consider it a simplifying assumption.

\subsection{Stable defects as excitations}

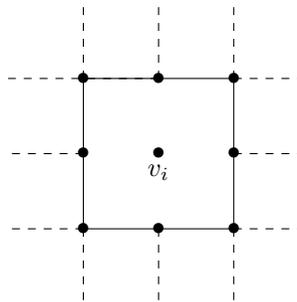
\begin{figure}[ht]
\centering
\begin{tikzpicture}[node distance=0.5cm]
	\node [black] at (0,1) {\textbullet};
	\node [black] at (1,1) {\textbullet};
	\node [black] at (2,1) {\textbullet};
	\node [black] at (0,0) {\textbullet};
	\node [black] at (1,0) {\textbullet};
	\node [black] at (2,0) {\textbullet};
	\node [black] at (0,-1) {\textbullet};
	\node [black] at (1,-1) {\textbullet};
	\node [black] at (2,-1) {\textbullet};

	\draw [dashed] (0,1) -- (0,2);
	\draw [dashed] (1,1) -- (-1,1);	
	\draw [-] (0,0) -- (0,1);
	\draw [dashed] (1,1) -- (1,2);
	\draw [-] (0,0) -- (0,-1);
	\draw [-] (0,1) -- (1,1);
	\draw [dashed] (2,1) -- (2,2);
	\draw [-] (2,1) -- (2,0);
	\draw [-] (2,0) -- (2,-1);
	\draw [-] (0,-1) -- (1,-1);
	\draw [-] (1,-1) -- (2,-1);
	\draw [dashed] (2,-1) -- (3,-1);
	\draw [-] (1,1) -- (2,1);
	\draw [dashed] (2,-1) -- (2,-2);
	\draw [dashed] (1,-1) -- (1,-2);
	\draw [dashed] (0,-1) -- (0,-2);
	\draw [dashed] (0,-1) -- (-1,-1);
	\draw [dashed] (0,0) -- (-1,0);
	\draw [dashed] (2,1) -- (3,1);
	\draw [dashed] (2,0) -- (3,0);	
	
	\node at (1,-0.25) {$v_i$};
\end{tikzpicture}
\caption{A section of the ground state of Eqs. \eqref{eqn:qmd_hamiltonian} or \eqref{eqn:trugenberger} with $\langle k \rangle=2d=4$. Depicted is an isolated defect surrounding vertex $v_i$. We chose $s_i=\ket{i,0}$, all other spins $s_j=\ket{j,1}$, but the key requirement is that $s_i$ is opposite to its neighbors.}
\label{fig:defect}
\end{figure}

We assume that the models described in Section \ref{sec:qmd} have been minimized to produce a regular ground state (the existence and precise nature of these ground states is extensively explored in \cite{trugenberger2015quantum,trugenberger2016random,tee2020dynamics} and we do not justify this here).
It was established in the original work that these ground states can have an excitation that possesses intriguing properties.
In Fig. \ref{fig:defect} we depict such an excitation where the uniform alignment of the spins, a characteristic of the Ising condensate phase of the graph, has been disturbed at $v_i$ by the spin being `flipped' relative to its neighbors.
In general at vertex $v_i$,  $s_{i} \neq s_{j}$, where $j$ ranges over all of the neighbors of $v_i$.
When the spins between neighbors anti-align, the presence of edges is disfavored, and the edges would intuitively be expected to be annihilated to  reduce the overall energy of the graph and result in a new stable minimum energy, which isolates the vertex and creates a topological `hole' in the graph.
This state has been analyzed using statistical mechanical arguments \cite{trugenberger2015quantum} and can be shown to be stable, even as the system is `cooled', and is thus a persistent feature once created.

Further analyzing the defect in Fig. \ref{fig:defect}, it is clear that the removal of the defect would require energy to re-establish links and flip the spin at $v_i$.
Effectively the defect stores the energy locally in the defect.
As such it embodies at least some of the properties of a particle; it is localized in space, has finite energy and is at least approximately conserved at least until enough energy is available to destroy the defect by creating edges and flipping the spin at $v_i$.
In the next section we will analyze the dynamics of such defects both in the continuum limit and also by exploring the evolution of the spins of vertices remote to the defect in the discrete case.

\section{Dynamics}
\label{sec:dynamics}

\subsection{The interaction Hamiltonian}
\label{ssec:meshdyn}

As discussed previously, our defects are persistent, and it can be shown \cite{tee2020dynamics} that they are eigenstates of the ground state Hamiltonian Eq. \eqref{eqn:qmd_hamiltonian}.
For them to become dynamic we must propose an interaction Hamiltonian that couples the defect localized at $v_i$ to other vertices in the graph.
Interactions should cause the defect to `close' at $v_i$ and create a corresponding defect surrounding  a different vertex $v_j$.
Such an interaction Hamiltonian should obey the following desiderata,

\begin{description}
    \item [Preserve the ground state] Any interaction Hamiltonian should have no coupling between vertices that have aligned spins, or are connected by an edge. This allows us to effectively separate analysis of the dynamics from the vacuum ground state.
    \item [Promote locality] Interactions between vertices close in the graph should be energetically more favorable than between distant ones. This locality is the origin of the approximate causality discussed later in the paper.
    \item [Self-consistently defined] Our proposed Hamiltonian should not require elements not already present in the existing model, and represent a minimal prescription for dynamics.
    \item [Dimensionally consistent] The final form for the Hamiltonian should have the correct dimensions of energy.
\end{description}

To address the first of these, using the Laplacian matrix and spin ladder operators for the vertices $\hat{s}^{\pm}_i$ we can construct a term such as $-\hat{s}^{+}_i ( 1+L_{ij} ) \hat{s}^{-}_j$,  the minus sign chosen to create an energy gradient that favors dynamics (recall that the eigenvalues of $\hat{s}_i^{\pm}$ are $1$).
The Laplacian matrix is the difference between the degree matrix $\Delta_{ij}$, which is the diagonal matrix defined as $\Delta_{ii}=k_i$, and the adjacency matrix.
As such the $(1+L_{ij})$ multiplier will be zero for any two nodes that are connected, and the combination of $\hat{s}^{\pm}_{i/j}$ will be zero whenever the two spins are identical.
To satisfy the third desideratum, we note that we can  express the Laplacian matrix in terms of edge annihilation/creation operators as follows:
\begin{equation}
	L_{ij} = \sum\limits_{k,j=0}^{k,j=N} \delta_{i}^{j}\hat{a}^{\dagger}_{ik} \hat{a}_{kj} - \hat{a}^{\dagger}_{ij} \hat{a}_{ij} \text{,}
\end{equation}

To promote locality we insert an inverse distance dependence on the interaction to cause the interaction to reduce as vertices become more distant.
In general this inverse polynomial could be of arbitrary degree.
In \cite{tee2020dynamics} we choose an inverse square, but for generality it could be an arbitrary integer value $\gamma > 1$.
We can define the graph distance between two vertices, $r_{ij}$ as,
\begin{equation}\label{eqn:hop_distance}
	r_{ij} = \sum\limits_{n=0}^{\infty} \left \{ \delta( A^n_{ij} ) \times \delta \left(\sum\limits_{p=0}^{p=n} A^p_{ij} \right) \right \}\text{.}
\end{equation}
and then insert this into the interaction as an $r_{ij}^{-\gamma}$ factor.
The $\delta( A^n_{ij} )$ and $\delta \left(\sum\limits_{p=0}^{p=n} A^p_{ij} \right)$ terms are to be understood as a delta function such that for integers $n \in \mathbb{Z}$, ~ $\delta(n \neq 0)=0$ and $\delta(0)=1$.

It remains to arrange  the dimensions of the interaction according to the last desideratum.
We set $\gamma=2$, which as the eigenvalues of the $\hat{s}^{\pm}_i$ operators are $\hbar/2$, we  must insert an energy in the denominator $\epsilon_m$ to balance the dimensions.
We interpret this energy term as the energy stored in a defect, as described above.
Using these definitions the proposal for the dynamic term of the Hamiltonian is

\begin{equation}\label{eqn:dynamic}
	\hat{H}_{int} = -\frac{gc^2}{2\epsilon_m r_{ij}^2} \hat{s}^{+}_i ( 1+L_{ij}) \hat{s}^{-}_j \text{.}
\end{equation}

This expression only deals with the spins at the vertices and does not take care of the edge states between them. 
To extend this requires some care in the analysis of the model.

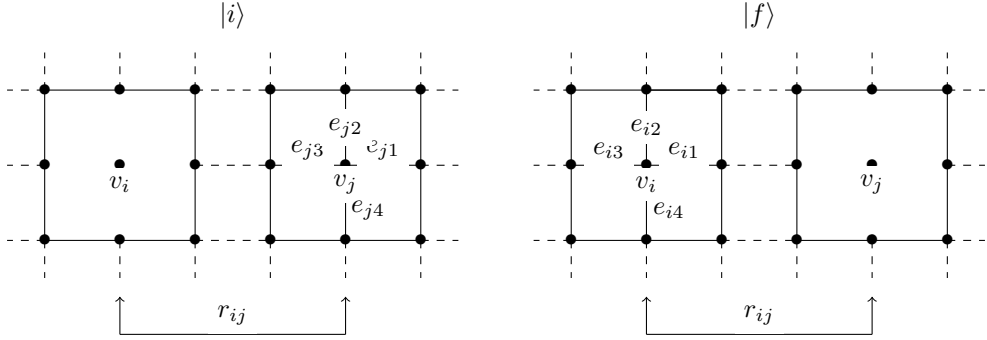
\begin{figure*}[http]
\centering
\begin{tikzpicture}[node distance=0.5cm]
	\node [black] at (0,1) {\textbullet};
	\node [black] at (1,1) {\textbullet};
	\node [black] at (2,1) {\textbullet};
	\node [black] at (0,0) {\textbullet};
	\node [black] at (1,0) {\textbullet};
	\node [black] at (2,0) {\textbullet};
	\node [black] at (0,-1) {\textbullet};
	\node [black] at (1,-1) {\textbullet};
	\node [black] at (2,-1) {\textbullet};

	\node [black] at (3,0) {\textbullet};
	\node [black] at (4,0) {\textbullet};
	\node [black] at (5,0) {\textbullet};
	\node [black] at (3,1) {\textbullet};
	\node [black] at (4,1) {\textbullet};
	\node [black] at (5,1) {\textbullet};
	\node [black] at (3,-1) {\textbullet};
	\node [black] at (4,-1) {\textbullet};
	\node [black] at (5,-1) {\textbullet};

	\draw [-] (0,0) -- (0,1);
	\draw [-] (0,0) -- (0,-1);
	\draw [-] (0,1) -- (1,1);
	\draw [-] (2,1) -- (2,0);
	\draw [-] (2,0) -- (2,-1);
	\draw [-] (0,-1) -- (1,-1);
	\draw [-] (1,-1) -- (2,-1);
	\draw [-] (1,1) -- (2,1);
	\draw [-] (3,1) -- (4,1);
	\draw [-] (3,0) -- (3,1);
	\draw [-] (3,0) -- (3,-1);
	\draw [-] (3,0) -- (4,0);
	\draw [-] (3,-1) -- (4,-1);
	
	\draw [dashed] (2,1) -- (3,1);
	\draw [dashed] (2,-1) -- (3,-1);
	\draw [dashed] (2,0) -- (3,0);
	\draw [dashed] (3,1) -- (3,1.5);	
	\draw [dashed] (2,1) -- (2,1.5);
	\draw [dashed] (1,1) -- (1,1.5);
	\draw [dashed] (0,1) -- (0,1.5);
	\draw [dashed] (0,-1) -- (0,-1.5);
	\draw [dashed] (3,-1) -- (3,-1.5);
	\draw [dashed] (2,-1) -- (2,-1.5);
	\draw [dashed] (1,-1) -- (1,-1.5);
	\draw [dashed] (0,1) -- (-0.5,1);
	\draw [dashed] (0,-1) -- (-0.5,-1);
	\draw [dashed] (0,0) -- (-0.5,0);
	\draw [dashed] (4,-1) -- (4,-1.5);
	\draw [dashed] (5,-1) -- (5,-1.5);
	\draw [dashed] (4,1) -- (4,1.5);
	\draw [dashed] (5,1) -- (5,1.5);
	\draw [dashed] (5,1) -- (5.5,1);
	\draw [dashed] (5,0) -- (5.5,0);
	\draw [dashed] (5,-1) -- (5.5,-1);
	
	\draw [-] (4,1) -- (5,1);
	\draw [-] (4,0) -- (5,0);
	\draw [-] (4,-1) -- (5,-1);
	
	\draw [-] (4,-1) -- (4,0);
	\draw [-] (4,0) -- (4,1);
	\draw [-] (5,-1) -- (5,0);
	\draw [-] (5,0) -- (5,1);
	
	\draw [-] (1,-2.25) -- (4,-2.25);
	\draw [->] (1,-2.25) -- (1,-1.75);
	\draw [->] (4,-2.25) -- (4,-1.75);
	\node[fill=white, text=black] at (2.5,-2) {$r_{ij}$};
		
	\node[fill=white, text=black] at (1,-0.25) {$v_i$};
	\node[fill=white, text=black] at (4,-0.25) {$v_j$};
	
	\node[fill=white, text=black] at (3.5,0.2) {$e_{j3}$};
	\node[fill=white, text=black] at (4.5,0.2) {$e_{j1}$};
	\node[fill=white, text=black] at (4,0.5) {$e_{j2}$};
	\node[ text=black] at (4.3,-0.6) {$e_{j4}$};

	\node [black] at (7,1) {\textbullet};
	\node [black] at (8,1) {\textbullet};
	\node [black] at (9,1) {\textbullet};
	\node [black] at (7,0) {\textbullet};
	\node [black] at (8,0) {\textbullet};
	\node [black] at (9,0) {\textbullet};
	\node [black] at (7,-1) {\textbullet};
	\node [black] at (8,-1) {\textbullet};
	\node [black] at (9,-1) {\textbullet};

	\node [black] at (10,0) {\textbullet};
	\node [black] at (11,0) {\textbullet};
	\node [black] at (12,0) {\textbullet};
	\node [black] at (10,1) {\textbullet};
	\node [black] at (11,1) {\textbullet};
	\node [black] at (12,1) {\textbullet};
	\node [black] at (10,-1) {\textbullet};
	\node [black] at (11,-1) {\textbullet};
	\node [black] at (12,-1) {\textbullet};
	
	\draw [-] (7,0) -- (7,1);
	\draw [-] (7,0) -- (8,0);
	\draw [-] (7,0) -- (7,-1);
	\draw [-] (7,1) -- (8,1);
	\draw [-] (8,0) -- (8,-1);
	\draw [-] (8,0) -- (8,1);
	\draw [-] (8,0) -- (9,0);
	\draw [-] (9,1) -- (8,1);
	\draw [-] (9,1) -- (9,0);
	\draw [-] (9,0) -- (9,-1);
	\draw [-] (7,-1) -- (8,-1);
	\draw [-] (8,-1) -- (9,-1);
	\draw [-] (8,1) -- (9,1);
	\draw [-] (10,1) -- (11,1);
	\draw [-] (10,0) -- (10,-1);
	\draw [-] (10,0) -- (10,1);
	\draw [-] (10,-1) -- (11,-1);
	\draw [-] (11,1) -- (12,1);
	\draw [-] (11,-1) -- (12,-1);
	\draw [-] (12,0) -- (12,1);
	\draw [-] (12,0) -- (12,-1);
	
	\draw [dashed] (9,1) -- (10,1);
	\draw [dashed] (9,-1) -- (10,-1);
	\draw [dashed] (9,0) -- (10,0);
	\draw [dashed] (10,1) -- (10,1.5);	
	\draw [dashed] (9,1) -- (9,1.5);
	\draw [dashed] (8,1) -- (8,1.5);
	\draw [dashed] (7,1) -- (7,1.5);
	\draw [dashed] (7,-1) -- (7,-1.5);
	\draw [dashed] (10,-1) -- (10,-1.5);
	\draw [dashed] (9,-1) -- (9,-1.5);
	\draw [dashed] (8,-1) -- (8,-1.5);
	\draw [dashed] (7,1) -- (6.5,1);
	\draw [dashed] (7,-1) -- (6.5,-1);
	\draw [dashed] (7,0) -- (6.5,0);
	\draw [dashed] (11,-1) -- (11,-1.5);
	\draw [dashed] (12,-1) -- (12,-1.5);
	\draw [dashed] (11,1) -- (11,1.5);
	\draw [dashed] (12,1) -- (12,1.5);
	\draw [dashed] (12,1) -- (12.5,1);
	\draw [dashed] (12,0) -- (12.5,0);
	\draw [dashed] (12,-1) -- (12.5,-1);

	\draw [-] (8,-2.25) -- (11,-2.25);
	\draw [->] (8,-2.25) -- (8,-1.75);
	\draw [->] (11,-2.25) -- (11,-1.75);
	\node[fill=white, text=black] at (9.5,-2) {$r_{ij}$};
		
	\node[fill=white, text=black] at (8,-0.25) {$v_i$};
	\node[fill=white, text=black] at (11,-0.25) {$v_j$};
	
	\node[fill=white, text=black] at (2.5,2) {$\ket{ i }$};
	\node[fill=white, text=black] at (9.5,2) {$\ket{ f }$};
	
	\node[fill=white, text=black] at (7.5,0.2) {$e_{i3}$};
	\node[fill=white, text=black] at (8.5,0.2) {$e_{i1}$};
	\node[fill=white, text=black] at (8,0.5) {$e_{i2}$};
	\node[ text=black] at (8.3,-0.6) {$e_{i4}$};

\end{tikzpicture}

\caption{A defect located at $v_i$ in a 2d graph and a target location in the graph at distance $r_{ij}$ centered around $v_j$. We consider the transition amplitude between a defined physical state $\ket{i}$ represented by the left hand section of the graph, and a final state $\ket{f}$ on the right hand side. To translate the vertex from $v_i$ to $v_j$, the marked edges $\{e_{j1}.\dots,e_{j4}\}$, must be destroyed and $\{e_{i1}.\dots,e_{i4}\}$ created. The defect at $v_i$ in $\ket{i}$ is localized surrounding $v_i$, and that in $\ket{f}$ around $v_j$. }
\label{fig:propagate}
\end{figure*}

In Fig. \ref{fig:propagate} we have an example of a transition in a $2$-dimensional graph that our interaction Hamiltonian should cause, including the creation and annihilation of the relevant edges. 
In the analysis that follows we will consider the case of a $2$-dimensional graph for simplicity, but any of the results are easily extended to higher dimensions.
The distinct physical states are characterized by the annihilation of the four edges surrounding $v_j$ and the creation of 4 new edges connecting $v_i$ to the boundary of the defect.
Both the ground state, and defects are valid eigenstates of the Hamiltonian $\hat{H}_{QMD}$, and so far $\hat{H}_{int}$ only affects the spins at the vertices.
We therefore need to alter the interaction Hamiltonian in such as way as the ground state is unaffected {\sl and} the defect is translated between $\ket{i}$ and $\ket{j}$.
We can diagrammatically represent our initial and final states, including the edges, using the following shorthand notation:
\begin{align}
	\ket{i} &= \ket{ \loopdef_{~i}~,  \vacedge_{~j} } \text{,} \\
	\ket{\scriptstyle f} &= \ket{ \vacedge_{~i}~, \loopdef_{~j} } \text{.}
\end{align}
This shorthand represents the states of the edges and spins surrounding the indicated vertex. The presence of a dot, indicates spin state of $\ket{1}$, the absence $\ket{0}$, and the presence of an edge between to dots $i$,$j$ the state $\ket{i,j,1}$ and $\ket{i,j,0}$ for its absence.
The vertex at the center of the diagram is indicated by the subscript.

To modify our Hamiltonian, let us define a new operator using a combinatorial sum of all possible subsets of vertices in the graph that could be neighborhoods of a vertex.
A neighborhood of a given vertex $v_i$ is defined as the collection of vertices in the graph that share an edge with $v_i$, and essentially we need to pick out the appropriate neighborhood of each of the vertices $v_i$, $v_j$ in $\ket{i}$,$\ket{f}$ and apply creation and annihilation operators to obtain the correct transition from $\ket{i} \rightarrow \ket{f}$.
On a general graph $G(V,E)$, we define a new operator $\hat{S}_i^{-}$,

\begin{equation}\label{eqn:S_minus}
	\hat{S}_i^{-} = \hat{s}_i^{-}  \sum\limits_{I \subset V} \prod\limits_{j \in I} \hat{a}_{ij} \text{,}
\end{equation}
where $I$ is any subset of vertices, and the sum ranges over every member of the power set of $V$,  that is every possible subset of $V$.
For an arbitrarily large graph this is clearly a combinatorially very large sum!

The sum over products creates a large number of combinations of $\hat{s}^{-}_i$ with edge annihilation operators, but fortunately as $\hat{a}_{ij} \ket{0} = 0$, most of the terms in the sum will vanish.
Diagrammatically we can represent the action of $\hat{S}_i^{-}$ on a fully connected state $\ket{\vacedge_{~i}}$ as:

\begin{equation*}
	\hat{S}_i^{-} \ket{ \vacedge_{~i}} = \ket{ \loopdef_{~i} } + 4 \times \ket{ \oneedge_{~i}} + 6 \times \ket{ \twoedge_{~i} } + 4 \times \ket{ \threeedge_{~i}} \text{,}
\end{equation*}
where the pre-factors arise from the symmetry and dimension of the graph, and count the possible equivalent number of states to the diagram in the ket.
Recall that the shorthand $\ket{\loopdef_{~i}}$ and $\ket{\vacedge_{~i}}$, is in fact the tensor product of every edge and vertex state (assumed to be uniformly $\ket{1}$ except at the considered vertices).
Each edge and spin has a set of basis vectors $\ket{0}$,$\ket{1}$, such that $\braket{n}{m}=\delta_{nm}$ up to a normalization constant.
This is helpful later when we take an expectation value of an operator as most of those terms will vanish due to orthonormality.
This simplification occurs because our initial and final states only contain $\vacedge$ and $\loopdef$ states and we note,
\begin{equation*}
	\bra{ \vacedge \text{, or } \loopdef}\ket{ \oneedge } = \bra{ \vacedge \text{, or } \loopdef}\ket{ \twoedge }= \bra{ \vacedge \text{, or } \loopdef}\ket{ \threeedge } = 0 \text{.}
\end{equation*}

We can define the equivalent inverse operation to $\hat{S}_i^{-}$ by taking the complex conjugate of Eq. \eqref{eqn:S_minus} as follows,

\begin{equation}\label{eqn:S+plus}
	\hat{S}_i^{+} = \left( \hat{S}_i^{-} \right )^{\dagger} =  \hat{s}_i^{+}  \sum\limits_{I \subset V} \prod\limits_{j \in I} \hat{a}_{ij}^{\dagger} \text{.}
\end{equation}
It is easy to verify that the operation of this expression on $\ket{\loopdef}$ has the desired result, producing again the following sequence of intermediate states,
\begin{equation*}
	\hat{S}_i^{+} \ket{ \loopdef_{~i}} = \ket{ \vacedge_{~i} } + 4 \times \ket{ \oneedge_{~i}} + 6 \times \ket{ \twoedge_{~i} } + 4 \times \ket{ \threeedge_{~i}} \text{.}
\end{equation*}

With these new operators that take care of both vertex and edge states, we can now redefine our dynamic Hamiltonian as,

\begin{equation}\label{eqn:H_mod}
	\hat{H}_{int}^{'} = -\frac{g}{\epsilon_m r^2_{ij} }\hat{S}_a^{+} ( 1 + L_{ij} ) \hat{S}_b^{-} \text{.}
\end{equation}
Because the eigenvalues of the creation and annihilation operators, like the spin ladder operators are identically $1$, the inclusion of the edge operators $\hat{S}_i^{\pm}$ does not change the sign or factors compared to Eq. \eqref{eqn:dynamic_simple} when operating on a state such as $\ket{ \loopdef_{~i}~,  \vacedge_{~j} }$, or $\ket{ \vacedge_{~i}~,  \loopdef_{~j} }$.

\subsection{Quantum mechanics and the continuum limit}
\label{ssec:waveeqn}
As a consistency check, we should recover regular non-relativistic quantum mechanics when we take the continuum limit of our model.
We define the continuum limit in this context as the process of shrinking the minimum distance (edge length) in the graph to zero.
To simplify the analysis we consider only the vertex states and ignore the edges, which we justify by observing that in the continuum limit the edges will disappear.

As the states as defined in Section \ref{sec:qmd} have no dependence upon time we are essentially in the Heisenberg picture.
The time dependence is then obtained by applying unitary time evolution and for a general increment in time $\tau$, the state vector $\ket{v_i, t}$ evolves as,
\begin{equation*}
	\ket{v_i, t+\tau} = e^{-i\hat{H}_{int} \tau/ \hbar} \ket{v_i,t} \mbox{.}
\end{equation*}
We compare this to the Taylor series expansion of the state in $\tau$,
\begin{equation*}
	\ket{v_i, t+\tau} = \ket{v_i,t} + \tau \frac{\partial \ket{v_i,t} }{\partial t}  + \order{\tau^2} \dots 
\end{equation*}
Expanding the exponential to $\order{\tau}$, and gathering terms we obtain,
\begin{equation}
	-\frac{\hbar}{i} \frac{\partial \ket{v_i,t} }{\partial t}  =  \frac{gc^2\hbar^2}{8\epsilon_m r_{ij}^2} \ket{v_i,t} +  \frac{gc^2\hbar^2}{8\epsilon_m r_{ij}^2}L_{ij} \ket{v_i,t} \text{.}
\end{equation}
In the last step, we assume that in the continuum limit $v_i$ is zero distance from $v_j$, and as $L_{ij}$ is just a number the $\hat{s}^{\pm}$ operators both act on the state $\ket{v_i,t}$, contributing $\hbar^2/4$.
The graph becomes a smooth manifold in the limit, and each vertex is thereby identified with a point in this $d$ dimensional space $\va*{x}$, and  $\ket{v_i,t} \rightarrow \ket{\va*{x},t}$.

Discrete dynamical systems on a graph often involve the Laplacian matrix, and in the continuum limit, the Laplacian matrix is equivalent to $-\nabla^2$, as discussed in \cite{chung1997spectral}.
As $r_{ij} \rightarrow 0$, we `renormalize' the coupling constant $g$ to absorb the resultant infinity. 
We recall that the edges of the graph are assumed to be a finite length (most often the Planck length $l_p$), and so the continuum limit is achieved by setting that distance $l_p=0$.
We reinterpret $g$ as the `bare' coupling constant, valid only when $l_p > 0$, and absorb the infinity into $g$, replacing it with the `physical' coupling constant $g_p$ defined as  $g_p=g_b /4 r_{ij}^2$.
Once we have made these substitutions we have,
\begin{equation}
	-\frac{\hbar}{i} \frac{\partial \ket{\va*{x},t} }{\partial t}  = \frac{g_p\hbar^2}{2m} \ket{\va*{x},t}  + \frac{g_p}{2m} \bigg( \frac{\hbar}{i} \nabla \bigg )^2\ket{\va*{x},t} \text{, }
\end{equation}
which we can immediately recognize as the Schr{\"o}dinger equation for a non-relativistic particle in a constant potential $V(\va*{x})=\frac{g_p\hbar^2}{2m}$.
This result, although seemingly remarkable, is simply the direct consequence of unitary evolution providing that the Laplacian matrix is present to recovered the $-\nabla^2$ term. 
It is the last point that is more significant, as the Laplacian matrix was inserted to guarantee consistency with the original Ising models, and indicates that the defects could be reasonable and self-consistent model of mass in the models.

\subsection{The time evolution of defects}

Having discussed how in the continuum limit one recovers non-relatavistic QM, lets us return to the time evolution of a defect in the graph.
Computing the time evolution of a defect in the discrete model will allow us in Section \ref{sec:simulations} to simulate the propagation of a defect in the ground state of an Ising model, and explore the causality of the interaction.

We continue to use as the concept of time the label we attach to the sequential evolution of the states of the spacetime graph, and there is no requirement for it to be continuous or discrete.
This label is globally the same for all observers in the graph and is therefore intrinsically  non-relativistic, occupying the same significance as in ordinary QM where it appears as a label on a quantum state without a corresponding operator and eigenstates.
In short we are not adding in `by hand' any causal structure or Lorentz invariance.

We start with a system at time $t_0$ where a single defect is present, centered at the vertex $v_i$.
To explore the dynamics we use the interaction defined by Eq. \eqref{eqn:H_mod} to compute the expected value of the spin $\hat{s}_j$ (assumed to be measured in the $z$ direction) at a vertex $v_j$ separated from a defect present at $v_i$ by distance $r_{ij}$.
By considering unitary evolution of the spin states using our interaction Hamiltonian we can calculate an expression for that expected value as a function of time, and we work in natural units where $\hbar=c=1$.
As the vertex and edge spin states are in the Heisenberg picture and independent of time, we evolve the initial state to time $t_0$, and the final state to time $t>t_0$ and  compute the following expected value,
\begin{equation}\label{eqn:expected_spin}
	\bra{f,t} \hat{s}_j \ket{i,t_0} = \bra{f} e^{i\hat{H}_{int}^{'} t} \hat{s}_j e^{-i\hat{H}_{int}^{'} t_0 }  \ket{i} \text{.}
\end{equation}
The unitary evolution should of course be undertaken using the whole Hamiltonian, including the appropriate ground state Hamiltonian Eq. \eqref{eqn:qmd_hamiltonian} or similar.
We can safely ignore that because both $\ket{f}$ and $\ket{i}$ are eigenstates of $H_{QMD}$, and due to orthonormality $\bra{f} H_{QMD} \ket{i} = \braket{f}{i}=0$.
Expanding the exponential and setting $t_0=0$, we have,
\begin{equation*}
	\bra{f} \hat{s}_j e^{i\hat{H}^{'}_{int} (t-t_0) }  \ket{i}= \bra{f}\ket{i} - \frac{igt}{2 \epsilon_m r_{ij}^2} \bra{f}\hat{s}_j \hat{S}_i^{+} \hat{S}_j^{-} \ket{i} +  \mathcal{O}(g^2) \text{.}
\end{equation*}
Noting that $\bra{f}\ket{i}=0$ and $\hat{s}_z \ket{i} = \hat{s}_z \ket{f} = \pm 1$.
In fact $\hat{s}_z$ and $\hat{S}^{\pm}$ anti-commute, but the negative sign that is picked up either from the value of $s_j$ in $\ket{f}$ or $\ket{i}$ after doubly commuting will not be relevant when we square the transition amplitude later.
The the only surviving term after expansion can therefore be represented using diagrammatic notation as,
\begin{equation*}
	\bra{ \vacedge_{~i},~ \loopdef_{j} } \hat{S}_i^{+} \hat{S}_j^{-} \ket{ \loopdef_{\substack
{i}},~ \vacedge_{~j} }. \text{.}
\end{equation*}
We can choose to operate the $\hat{S}^{+}_i$ operator on the left or right, but choosing the left we have,
\begin{align*}
	&\bra{ \vacedge_{~i},~ \loopdef_{j} } \hat{S}_{\substack{ i \\ \leftarrow}}^{+}   \rightarrow \bra{ \vacedge_{~i},~ \loopdef_{j}}  \left( \hat{S}^{+}_i \right)^{\dagger} = \\
	&\bra{ \vacedge_{~i},~ \loopdef_{j} } \hat{S}_i^{-} = \bra{ \loopdef_{~i},~ \loopdef_{~j} } + 4\times \bra{ \oneedge_{~i},~ \loopdef_{~j} }\\
	&+ 6 \times \bra{ \twoedge_{~i},~ \loopdef_{~j} } + 4 \times \bra{ \threeedge_{~i},~ \loopdef_{~j} } \text{.}
\end{align*}
Similarly for the other operator we have,
\begin{align*}
	\hat{S}^{-}_j \ket{ \loopdef_{~i},~ \vacedge_{~j} } =  &\ket{ \loopdef_{~i},~ \loopdef_{~j} } + 4\times \ket{ \loopdef_{~i},~ \oneedge_j }\\
	+ 6 \times & \ket{ \loopdef_{~i},~ \twoedge_j } + 4 \times \ket{ \loopdef_{~i},~ \threeedge_j } \text{.}
\end{align*}

When we combine the outcome of operating on the initial and final states with $\hat{S}^{\pm}$  only the terms in $\bra{ \loopdef_{~i},~ \loopdef_{~j}} \ket{ \loopdef_{~i},~ \loopdef_{~j}}$ survive, as all other states are orthogonal.
This is because our simplified notation hides the fact that these states are in fact the tensor product states across every vertex and edge state in the system. 
Any non-identical component state in $\braket{f}{i}$ contributes a multiplicative zero eliminating such a term.

Having understood that the edges states do not complicate the calculation we can simplify the computation and focus on the spin states of the vertices $v_i$ and $v_j$.
As the two points are disconnected,  we  simplify Eq. \eqref{eqn:dynamic} for the interaction between the two vertices to,
\begin{equation}\label{eqn:dynamic_simple}
	\hat{H}_{int} = \alpha  \hat{s}^{+}_i  \hat{s}^{-}_j \text{, with } \alpha= -\frac{g}{2\epsilon_m r_{ij}^2} \text{.}
\end{equation}

To compute Eq. \eqref{eqn:expected_spin}, we evolve our states using the interaction Hamiltonian as before.
We can write our initial and final states (retaining $t_0$ for now) as,
\begin{align}
	\ket{i,t_0}   &= e^{-i \hat{H}_{int} t_0} \ket{i} \text{,} \label{eqn:initial_t}\\
	\ket{f,t}     &= e^{-i \hat{H}_{int} t } \ket{f} \text{.} \label{eqn:final_t}
\end{align}

As with the edge states this is highly economical notation as in fact these are states of the whole graph and are properly defined in the full tensor product Hilbert space.
We introduce the simplification of considering only pairwise interactions of the vertices and  work in the restricted Hilbert space spanned by $\mathcal{H}_i \otimes \mathcal{H}_j$, that is the tensor product of the Hilbert space of the quantum spin states at $v_i$ and $v_j$.
At each of the vertices, we have a simple set of basis vectors, which we can chose to define the Hilbert space as $\mathcal{H}_i=\text{span}\{ \left( \begin{smallmatrix}1\\0\end{smallmatrix} \right ), \left ( \begin{smallmatrix}0\\1\end{smallmatrix} \right ) \}$.
In this representation our spin operators are,
\begin{align}
		\hat{s}^+ &= \begin{pmatrix} 0 & 1 \\ 0 & 0 \end{pmatrix} \text{,} \hat{s}^- = \begin{pmatrix} 0 & 0 \\ 1 & 0 \end{pmatrix} \text{,}\\
		\hat{s}_{i} &= \begin{pmatrix} 1 & 0 \\ 0 & -1 \end{pmatrix} \text{,}
\end{align}
which satisfy the normal anti-commutation relationships $\{ \hat{s}_z , \hat{s}^{\pm} \} = 0, [ \hat{s}_z , \hat{s}^{\pm} ] = \pm 2 \hat{s}^{\pm}$.

To avoid writing out the representation in matrix form for the tensor product space, we write in abbreviated form the initial and final states $\ket{i} = \ket{ 0,1}$, $\ket{f} = \ket{1,0}$, to indicate the state in the Hilbert space $\mathcal{H}_i \otimes \mathcal{H}_j$ corresponding to $v_i$ having spin down $v_i$ spin up in the initial state, with the opposite values in the final state.

The spin operators anticommute and obey a similar exponential expansion to c-numbers obeying a Grassmann algebra \cite{ryder1996quantum}.
This simplifies the exponential in the time evolution operator, and we can demonstrate this when we expand it as follows,
\begin{equation*}
	e^{-i \hat{H}_{int} t}  = 1 - i \alpha t \hat{s}_i^+  \hat{s}_j^-  + \left ( i \alpha t  \right)^2 \hat{s}_i^+  \hat{s}_j^- \hat{s}_i^+  \hat{s}_j^-  + \mathcal{O}(g^3) \dots 
\end{equation*}

For powers higher than $\mathcal{O}(g)$, this involves further pairs of $\hat{s}_i^+  \hat{s}_j^- $ operators.
By definition $\hat{s}_i^+  \hat{s}_j^- \ket{0,1} = \ket{1,0}$, and $\hat{s}_i^+  \hat{s}_j^- \ket{1,0} = 0$, so terms $\mathcal{O}(g^2)$ and above will not contribute to the expansion.
We conclude that,
\begin{equation}
	e^{-i \hat{H}_{int} t}  = 1 - i\alpha t \hat{s}_i^+  \hat{s}_j^-  \text{  ,}
\end{equation}
which is the standard form of the exponential function of anticommuting operators.

Setting $t_0 = 0$ and operating on the right with the remaining $e^{i \hat{H}_{int} t}$ term, we have,
\begin{align*}
	e^{i \hat{H}_{int} t} \ket{0,1} &= \left (\hat{ \unit} + i \alpha t \hat{s}_i^+  \hat{s}_j^-  \right ) \ket{0,1} \text{,}\\
					      &= \ket{0,1} + i \alpha t \hat{s}_i^+  \hat{s}_j^-  \ket{ 0,1 } \text{.}\\
					      & \text{ Noting that $\hat{s}_i^+  \hat{s}_j^-  \ket{ 0,1 } =  \ket{1,0}$ } \\
					      &= \ket{0,1} + i \alpha t \ket{ 1,0 } \text{.}
\end{align*}

We now multiply on the left by $\bra{ 1,0 }\hat{s}_j$, and as $	\hat{s}_j	\ket{ 1,0} = - \ket{1,0}$, and obtain our final result,
\begin{equation}\label{eqn:tran_result}
	\bra{f,t} \hat{s}_i \ket{i,t_0=0} = -i \alpha t \text{.}
\end{equation}

If we add in the edge states and evolve forward using the full interaction Eq. \eqref{eqn:dynamic} we obtain the same result as the edge states simply contribute $\pm 1$ to the term, which when we square for a transition probability are irrelevant.

The transition amplitude is now in a form we can test with simulations in Section \ref{sec:simulations}.
Essentially this allows us to determine the probability of a defect interacting with a remote vertex at a distance $r_{ij}$ after time $t$.
Strict causality would demand $\bra{f,t} \hat{s}_i \ket{i,t_0=0}^2$ decays rapidly to $0$ when $r_{ij} > ct$, but as we have no causal structure in our model, perhaps it should be expected that there is no such constraint on the expectation value. 
The presence of both time and distance though in Eq. \eqref{eqn:tran_result} suggests that the situation in our model may be more complex.
There is a more rigorous way to treat causality in lattice bound quantum systems that focuses on the commutator of interactions, known as Lieb-Robinson bounds that we turn to next.

\subsection{Lieb-Robinson bounds and causality}
\label{ssec:lieb}
Discrete lattice systems with long range interactions have been understood to exhibit bounds on the propagation speeds of such interactions, independent of any imposed relativistic causality \cite{lieb1972finite}.
This upper bound on the propagation speed is commonly referred to as a Lieb-Robinson (LR) bound.
Our interaction Hamiltonian has many of the characteristics of those that exhibit a LR bound, including a long range interaction that falls in strength according to a power law \cite{tran2020hierarchy}.
The computation of the LR bound involves estimating the operator norm of the commutator of arbitrary observables at physically separated sites in the lattice.
It can be proven that in certain circumstances this commutator bound is such that it decays away exponentially to zero for distances greater than a value that increases with time.
If the commutator of arbitrary observables is zero between two points in the lattice, they cannot be causally connected, and hence the supposition of a maximum propagation velocity.
If such a bound existed in our model, even approximately, we would have an indication of the presence of a causal structure intrinsic to the emerged geometry.

The calculation  of a LR bound is highly technical, and  we refer the reader to the literature \cite{lieb1972finite,sims2011lieb,nachtergaele2006propagation,nachtergaele2007lieb} for more details.
In essence, we state it in terms of two collections of non overlapping lattice points containing two vertices $v_i$ and $v_j$, separated by a distance $r_{ij}$ in the lattice, where $r_{ij}$ is the distance determined by an appropriate graph metric such as minimum edge traversal distance.
The definition of vertex separation defined by  Eq. \eqref{eqn:hop_distance}, for example, would be a valid distance measure.

Denoting $\norm{\cdot}$ as the operator bound of an operator we state the main result of LR as follows,
\begin{equation}\label{eqn:lr_bound}
    \norm{  \comm{\tau_t(\hat{A}_X)}{\hat{B}_Y } } \leq \frac{ \norm{\hat{A}} \norm{\hat{B}} }{C_a} \exp \left ( 2 \norm{\Phi}_a C_a|t| -1 \right ) D(X,Y) \text{.}
\end{equation}
The statement bounds the commutator between any two well behaved operators representing observables $\hat{A}$ and $\hat{B}$, whose supports are the disjoint sets $X$ and $Y$ considered as subsets of a $d$ dimensional lattice $\mathbb{Z}^n$, separated by a distance function $D(X,Y)$.
Between the sets $X,Y$, there is an interaction $\Phi_{ij}$ defined between any pair of lattice points such as $v_i \in X$ and $v_j \in Y$.
The interaction is assumed to be approximately short range in that it is possible to express a bound upon the interaction in terms of the distance function $D(X,Y)$ and a constant $C_a$, the convolution constant, that is defined by examining the geometry of the lattice upon which the system is defined. 
A modified distance function is also defined $D_a(r_{ij})=e^{-ar_{ij}} D(r_{ij}), a > 0$, and we note that $D_a(r_{ij}) < D(r_{ij})$.
Specifically, the distance function must satisfy the inequality,
\begin{equation}
    \sum\limits_{v_z} D_a(r_{iz}) D_a(r_{zj}) \leq C_a D_a(r_{ij}) \text{,} 
\end{equation}
for all points in the lattice $v_z \neq v_i,v_j$
The constant $a$ is the subscript referred to as the convolution constant $C_a$.
With this new distance function we define a new operator norm for $\Phi_{ij}$ as,
\begin{equation*}
    \norm{\Phi_{ij}}_a=\sup\limits_{i,j} \sum\limits_{i,j} \frac{\norm{ \Phi_{ij}  }}{D_a(r_{ij})}
\end{equation*}
The bound expresses approximate causality  because via the modified distance function the exponential term in Eq. \eqref{eqn:lr_bound} becomes $\exp \left ( 2 \norm{\Phi}_a C_a|t| -ar_{ij} \right )$, which we note decays quickly to zero for $r_{ij} > v_{LR}t$, where $v_{LR}$, the Lieb-Robinson velocity, is defined as
\begin{equation}
    v_{LR} = \frac{ 2 \norm{\Phi_{ij}}_a C_a }{a}\text{.}
\end{equation}

To make use of this calculation for our model, we need to be more precise about the calculation of $C_a$, $a$, $D(X,Y)$ and $r_{ij}$, and re-cast our dynamical Hamiltonian so it can be written as a LR interaction $\Phi$.
Without computing a precise result for the LR bound of our interaction, we present an overview of how the  calculation could be performed, and an argument for the existence of a bound.

We consider our interaction $\hat{H}_{ij}=\alpha \hat{s}_i^{+} \hat{s}_j^{-}$ to be split into a spin-spin and distance function, which we slightly modify to $\hat{H}_{ij}=\beta (1+r_{ij})^{-2} \hat{s}_i^{+} \hat{s}_j^{-}$, with $\beta=-\frac{g}{2\epsilon_m}$.
The change $(1+r_{ij})^{-2}$ is to ensure that our interaction does not become infinite for $r_{ij} = 0$, as required by the standard set up of the LR bound.
If we define $\Phi_{ij}=\beta\hat{s}_i^{+} \hat{s}_j^{-}$, the existence of the LR bound is thus determined by the norm of this interaction.

To determine whether this exists we note that $\Phi_{ij}^{\dagger}= \Phi_{ji}$, where although we have omitted the operator notation on $\Phi_{ij}$ it should be remembered that it is an operator.
From elementary operator theory $\norm{ \Phi_{ij}}$ is the square root of the maximum eigenvalue of $\Phi_{ij}^{\dagger} \Phi_{ij}=\Phi_{ji} \Phi_{ij}$.
Because of the order of spin ladder operators, we have only to consider the state $\ket{\loopdef_{~i},~ \vacedge_{~j} }$, for arbitrary vertex $v_j$, as the operation of $\Phi_{ji} \Phi_{ij}$ on any other state is zero.
One can straightforwardly verify that,
\begin{equation*}
    \Phi_{ji} \Phi_{ij} \ket{\loopdef_{~i},~ \vacedge_{~j} } = \beta^2 \ket{\loopdef_{~i},~ \vacedge_{~j} } \text{,}
\end{equation*}
and so the norm exists, and we state the result $\norm{\Phi_{ij}}=\beta$.

This argument is neither a calculation of, nor a precise proof of the existence of the LR bound, but does indicate that such a result may be obtainable. 
A more rigorous treatment requires the computation of $C_a$, which arises from the geometry of the graph, and the value of $\norm{\Phi_{ij}}_a$, and is the subject of ongoing work.
We can, however, make progress by investigating the propagation of our defects according to the interaction Hamiltonian Eq. \eqref{eqn:dynamic_simple} using numerical simulations in the following section.

\section{Simulations}

We simulate our interaction using an idealized ground state in a $2$-dimensional regular graph.
In previous work it has been argued that such a graph is a ground state of Ising emergent geometry models \cite{trugenberger2015quantum,tee2020dynamics}, and so it is a reasonable simplification.
At the center of this graph we create a single defect at time $t=0$.
We then evolve forward the model using the results from Section \ref{sec:dynamics}.
For every vertex at a time $t$ in the graph we can compute the expected value of the spin using Eq. \eqref{eqn:tran_result}, so that,
\begin{equation}
    \expval{\hat{s}_j(t)} =  | \bra{\psi_{\scriptscriptstyle f}, t} \hat{H}_{int} \ket{\psi_i, t=0} |^2 \hat{s}_j(t=0) \text{,}
\end{equation}
where $\ket{\psi_{\scriptscriptstyle f}} = \ket{\vacedge_{~i},~ \loopdef_{~j}}$ and $\ket{\psi_i}=\ket{\loopdef_{~i},~  \vacedge_{~j}}$, and we induce the time dependence by evolving forward the states using Eqs. \eqref{eqn:initial_t} and \eqref{eqn:final_t}.

\begin{figure*}[ht]
    \centering
	\begin{subfigure}[t]{0.45\textwidth}
		\centering
		\includegraphics[scale=0.44]{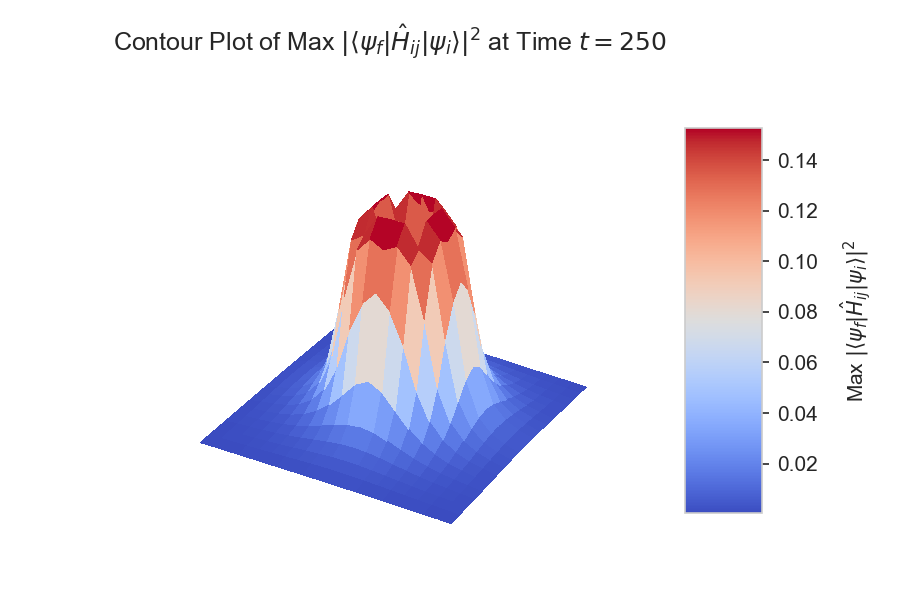}
		\caption{Simulation of $\abs{ \bra{\psi_j}\Hat{H}_{ij}\ket{\psi_i} }^2$ arrested at time $t=250$, with $v_j$ located at origin.}
		\label{fig:Prop250}	
	\end{subfigure}%
	~ 
	\begin{subfigure}[t]{0.45\textwidth}
		\centering
		\includegraphics[scale=0.44]{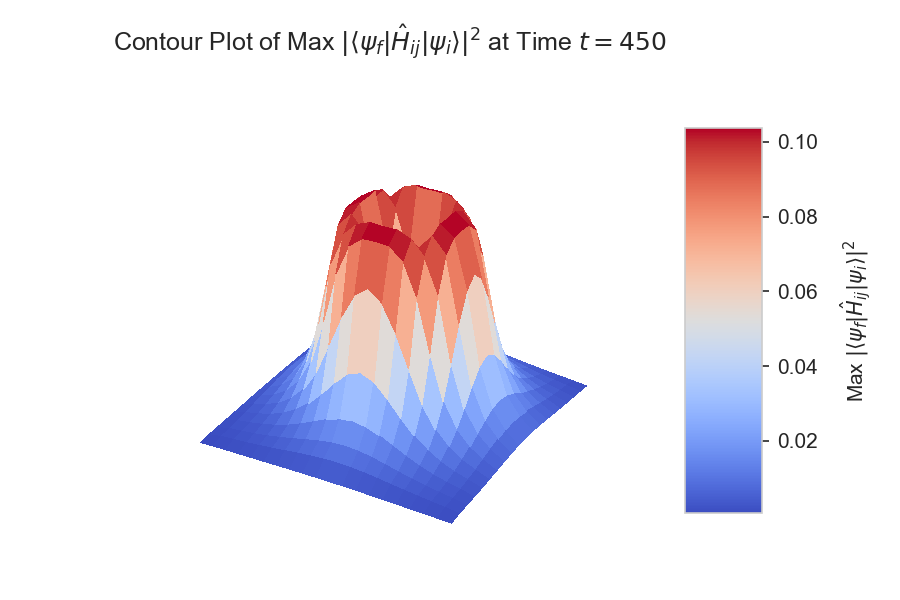}
		\caption{Simulation of $\abs{ \bra{\psi_j}\Hat{H}_{ij}\ket{\psi_i} }^2$ arrested at time $t=450$, with $v_j$ located at origin.}
		\label{fig:Prop450}
	\end{subfigure}
	~
	\begin{subfigure}[t]{0.45\textwidth}
		\centering
		\includegraphics[scale=0.44]{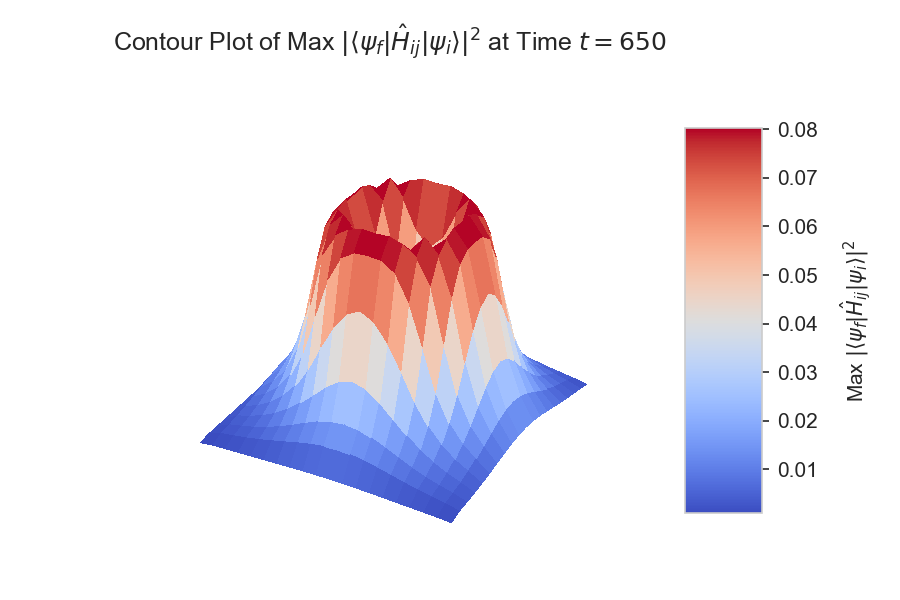}
		\caption{Simulation of $\abs{ \bra{\psi_j}\Hat{H}_{ij}\ket{\psi_i} }^2$ arrested at time $t=650$, with $v_j$ located at origin.}
		\label{fig:Prop650}	
	\end{subfigure}%
	~ 
	\begin{subfigure}[t]{0.45\textwidth}
		\centering
		\includegraphics[scale=0.44]{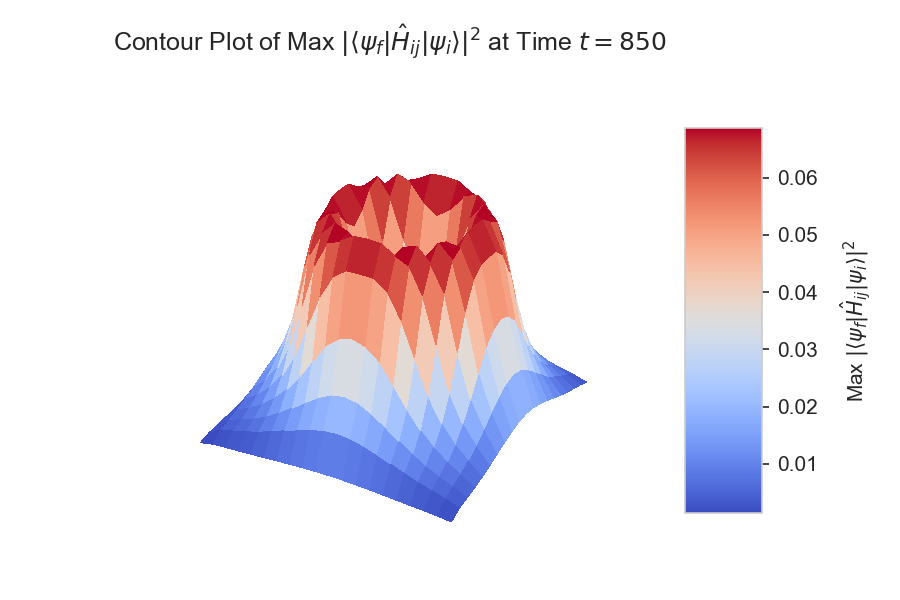}
		\caption{Simulation of $\abs{ \bra{\psi_j}\Hat{H}_{ij}\ket{\psi_i} }^2$ arrested at time $t=850$, with $v_j$ located at origin.}
		\label{fig:Prop850}
	\end{subfigure}
	\caption{Starting with a defect instantaneously created at the origin at time $t=0$, we present a numerical simulation of the amplitude of the interaction Hamiltonian Eq. \eqref{eqn:tran_result} at progressively later times. The model has implicit circular symmetry and we observe that the maximum amplitude of the interaction propagates outwards from the center. Either side of the wave front the interaction decays rapidly.}
	\label{fig:DefectPropagation}
\end{figure*}

In Fig. \ref{fig:DefectPropagation} we arrest the simulation at times $t=250,450,650$ and $850$ and plot the value of this expectation value.
We can see in the plots the presence of a distinct maximum that appears to propagate away from the defect as time increases. 
It is well localized in the graph, in the sense that the maximum decays away quickly for values of $r_{ij}$ less than or greater than the position of the maximum.
We could interpret this to indicate that the interaction at a given value of $t$, {\sl does not} interact with lattice sites further away than the position of the maximum of $\expval{\hat{s}_j(t)}$, which we call $r_m (t)$ to underline that it is time dependent.
For earlier times $t^{'}<t$ the maximum of the interaction will have passed through all points closer to the defect than $r_m(t)$, but will not have interacted significantly with more distant vertices.

\label{sec:simulations}

\begin{figure*}[ht]
	\centering
	\begin{subfigure}[t]{0.45\textwidth}
		\centering
		\includegraphics[scale=0.44]{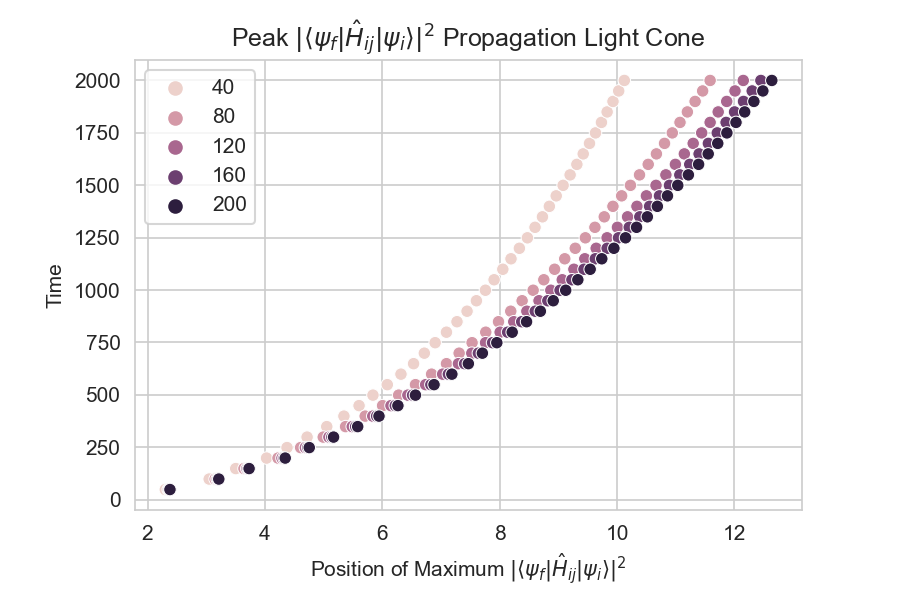}
		\caption{Starting at time $t=0$ with an instantaneously created defect at the origin, we compute the position of the maximum of $\abs{ \bra{\psi_j}\Hat{H}_{ij}\ket{\psi_i} }^2$ for a $2$-dimensional graphs of sizes up to $N=200$. }
		\label{fig:LightCone}	
	\end{subfigure}%
	~ 
	\begin{subfigure}[t]{0.45\textwidth}
		\centering
		\includegraphics[scale=0.44]{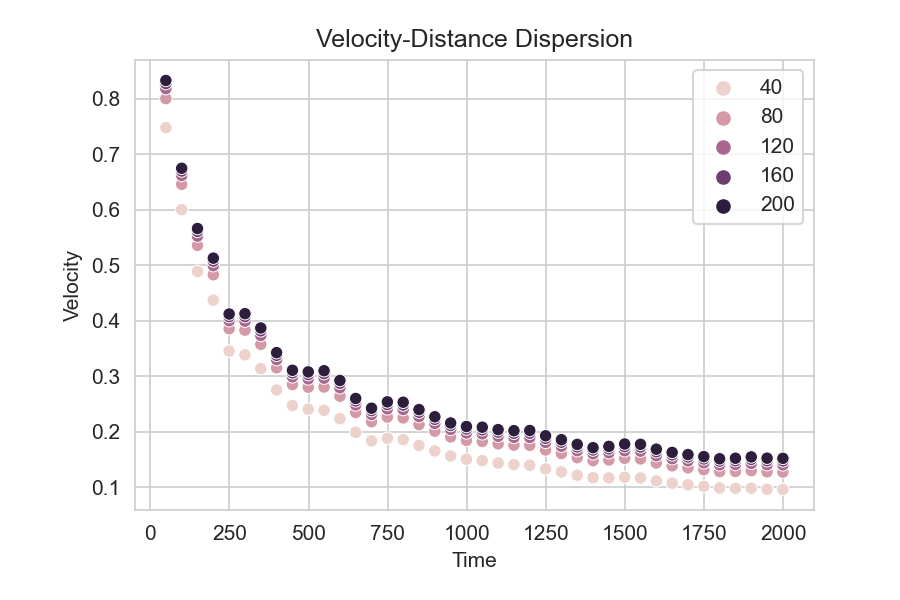}
		\caption{For the same defect created instantaneously at time $t=0$, we compute the velocity of propagation of the maximum of $\abs{ \bra{\psi_j}\Hat{H}_{ij}\ket{\psi_i} }^2$ away from the origin. This is performed for $2$-dimensional graphs of sizes up to $N=w00$.}
		\label{fig:VelDispersion}
	\end{subfigure}
	\caption{For a defect created instantaneously at the origin at time $t=0$, we compute light-cones and velocity dispersion diagrams for increasingly large graphs. It is visible as we scale the graph up that the propagation of the defect according to Eq. \eqref{eqn:tran_result} is tending to a limit. The drop off of velocity and change in the light cone for greater distances away from the origin is an effect of the finite size of the graph.}
	\label{fig:Causality}
\end{figure*}

We can analyze the the evolution of $r_m$ with time numerically and in Fig. \ref{fig:Causality} we present two graphs that show this behavior.
For computational efficiency, our simulation is conducted on a finite size graph, and as $r_m$ gets close to the boundary of the graph we encounter the effect of it.
In Fig. \ref{fig:VelDispersion} we plot the speed with which the interaction propagates away from the defect, $\partial_t r_m(t)$, for a variety of graphs of size $N=40$ to $N=200$.
As the graph increases in size the velocity overall increases, whilst appearing to converge to a maximum value at each distance, but the most distinct feature is the reduction in $\partial_t r_m(t)$ for larger $t$.
This is to be expected, and is once again interpreted as the effect of the finite size of the graph.

Perhaps more interesting is Fig. \ref{fig:LightCone}, where we plot the position of $r_m(t)$ of the interaction against time.
As the size of the graph increases the relation tends to a limiting curve, and for small $r_m(t)$ the relationship is approximately linear.
We believe that this could indicate the emergence of an approximate light-cone and the presence of a Lieb-Robinson bound for the system.
Of course the position $r_m(t)$ is not exact as we are dealing with a quantum amplitude which is non zero for $r \neq r_m(t)$.
Nevertheless this is a motivational result, although short of a direct comparison with an exact analytical computation of a LR bound.

\section{Conclusion}
\label{sec:conclusion}
Our principle focus in this work was to investigate the dynamics of a defect in an emerged geometry obtained as the ground state of Eq. \eqref{eqn:qmd_hamiltonian}.
We have re-examined the result previously presented in \cite{tee2020dynamics} and provided further investigation of the dynamics including a more rigorous treatment of the edge states in the graph.

Further, we speculated that the interaction Hamiltonian Eq. \eqref{eqn:dynamic} may carry with it an approximate causality in the form of a Lieb-Robinson bound. 
Numerical simulations provided evidence that this may be the case, but it is an open question to determine analytically whether our interaction supports such a bound, and indeed if it is consistent with the numerical experimentation. 

The role of time, and a causal structure is a key question in the combinatorial quantum gravity program.
We hope that this more detailed analysis of dynamical extensions to the Ising models provides an indication of how time could be incorporated into such models, simply by extending to include a minimal prescription for defect interactions.
It is possible that the subtle interplay between geometrical emergence and quantum mechanics could in these settings also provide an origin of emergent time and causality.

\section*{Acknowledgements}
I would like to thank the many fruitful conversations and debates I have had with George Zahariade and Paul Davies of the Beyond Institute at ASU. I would also like to note the helpful hints from Bruno Nachtergaele on the complex subject of Lieb-Robinson bounds.

\bibliographystyle{unsrt}
\bibliography{DefectDynamics}

\end{document}